\documentclass[prd,showpacs,amsmath,amssymb,12pt]{revtex4}
\usepackage{latexsym}
\usepackage{bm}

\def\k{\kappa} 
\def\g{\gamma} 
\def\a{\alpha} 
\def\b{\beta}

\def\k{\kappa}
 
\def\m{\mu} 
\def\n{\nu}

\def\s{\sigma}

 \def\O{\Omega}
\def\be{\begin{equation}}
\def\ee{\end{equation}}

\begin{document}

\title{Spherically symmetric solutions of Einstein + non-polynomial
gravities}

\author{S. Deser}
\email{deser@brandeis.edu}
\affiliation{California Institute of Technology, Pasadena, CA 91125 and \\
Department of Physics, Brandeis University, Waltham, MA 02254, U.S.A.}

\author{{\" O}zg{\" u}r Sar{\i}o\u{g}lu}  
\email{sarioglu@metu.edu.tr}
\affiliation{Department of Physics, Faculty of Arts and  Sciences,\\
             Middle East Technical University, 06531, Ankara, Turkey}

\author{Bayram Tekin}  
\email{btekin@metu.edu.tr}
\affiliation{Department of Physics, Faculty of Arts and  Sciences,\\
             Middle East Technical University, 06531, Ankara, Turkey}

\date{\today}

\begin{abstract}
We obtain the static spherically symmetric solutions of a class of 
gravitational models whose additions to the General Relativity (GR) action 
forbid Ricci-flat, in particular, Schwarzschild geometries. These theories
are selected to maintain the (first) derivative order of the
Einstein equations in Schwarzschild gauge. Generically, the
solutions exhibit both horizons and a singularity at the origin, except 
for one model that forbids spherical symmetry altogether. Extensions to 
arbitrary dimension with a cosmological constant, Maxwell source and 
Gauss-Bonnet terms are also considered.
\end{abstract}


\maketitle

\section{\label{intro} Introduction}
To better appreciate the nature of Ricci-flat geometries, particularly 
the fundamental exterior Schwarzschild solution of GR, it is useful to 
explore some alternatives, keeping as much as possible of the GR physics. 
The study of alternate theories is of course an enormous industry, but 
our interest here is to probe as simply as possible how the Schwarzschild 
geometry is deformed by additional terms. This will give a little more 
insight into the necessity for and variety of horizons, on which one 
could test the universality of our present black hole ideas. To be sure, 
we will only look under ``lampposts'' -- models for which spherically 
symmetric solutions can be found explicitly. Our deviations from GR 
consist of those terms, non-polynomial in the Weyl tensor, whose virtue 
is to preserve the (first) derivative order of the GR equations in 
Schwarzschild gauge and to provide the simplest non-Ricci flat extensions
of GR.

Our solution technique (introduced by Weyl \cite{W} for pure GR but
justified later \cite{P,DT}) is to insert in the action a gauge fixed 
metric already endowed with the desired symmetries. Gauge fixing means 
only that the Bianchi identities become implicit, while ansatzing a 
spherically symmetric $g_{\m\n}$ will enormously simplify the labor of 
obtaining the field equations for the two functions on which it depends. 
We are assured by \cite{P} that all solutions are obtained thereby. We 
will also include a cosmological constant, a Maxwell source and -- in 
$D>4$ -- Gauss-Bonnet terms.

\section{\label{body} The models}

In this section, we keep to $D=4$ for ease of notation. The most general 
spherically symmetric metric in Schwarzschild coordinates is usefully
written as
\be
ds^2 = - a(r,t) \, b^2(r,t) \, dt^2 + \frac{dr^2}{a(r,t)} + r^2 \, d\O_{2} \;.
\label{met}
\ee
All nonvanishing components of the mixed Weyl tensor, 
\( C_{\m\n}~^{\a\b} \), are proportional to the single function $X$
\be 
X(r,t) \equiv \frac{1}{r^2} (2 (a - 1) - 2 r a^{\prime} 
+ r^2 a^{\prime\prime}) + \frac{1}{r b} (3 r a^{\prime} b^{\prime} - 2 a 
(b^{\prime}- r b^{\prime\prime})) + \frac{1}{b} \partial_{t} 
\Big( \frac{1}{a^2 b} \partial_{t} a \Big) \,;
\ee
here primes denote radial derivatives. This means that any scalar of order
$n$ in the Weyl tensor $C$ is proportional to $X^{n}$. Indeed, this fact
is part of a general classification \cite{DR} of all algebraic curvature
invariants of spherical geometries. This classification also informs us that
the local actions that maintain the derivative order of the GR equations
for the metric (\ref{met}) are non-polynomial terms of the form 
\( (\mbox{tr} \, C^n)^{1/n} \), \( (\mbox{det} \, C^n)^{1/n} \), etc. 
Its Ricci scalar is 
\be 
R = -\frac{1}{r^2} \Big( 2 (a-1) + 4 r a^{\prime} 
+ r^2 a^{\prime\prime} \Big) - \frac{1}{r b} (3 r a^{\prime} b^{\prime}
+ 2 a (2 b^{\prime} + r b^{\prime\prime}) ) 
+ \frac{1}{b} \partial_{t} \Big( \frac{1}{a^2 b} \partial_{t} a \Big) \, . 
\ee
The actions we consider then are, in units of $\k=1$,
\be 
I = \frac{1}{2} \int d^4 x \, \sqrt{-g} \, \Big( R + 
\b_{n} \, | \mbox{tr} \, C^n |^{1/n} \Big) \, , \label{act}
\ee
where
\be 
\mbox{tr} \, C^n \equiv C_{ab}~^{cd} C_{cd}~^{ef} \dots 
C_{..}~^{pq} C_{pq}~^{ab} = \Big(- \frac{1}{3} \Big)^{n} \; 
\Big[ 2 + (-2)^{2-n} \Big] \, X^{n} \, ,
\ee
for $n$ copies of the Weyl tensor $C$, and there can be any number of
such Weyl additions. Without loss of generality, we consider $n=2$.
Defining $\s = \b_{2}/\sqrt{3}$, the action (\ref{act}), up to boundary 
terms, reduces to the almost trivial form
\be 
I \to \int_{0}^{\infty} dr \Big[ (1 - \s) (a r b^{\prime} + b) + 3 \s a b 
\Big] \, . \label{ract}
\ee 
Note that all the time-derivative terms have dropped out of the action.
[While this does not in itself guarantee Birkhoff's theorem, the methods
of \cite{DF} should do so.] The resulting two field equations decouple 
and give the immediate solution
\be 
a(r) = \frac{1 - \s}{1 - 4 \s} + a_1 \, r^{(4 \s - 1)/(1 - \s)} 
\, , \qquad b(r) = b_1 \, r^{3 \s / (\s - 1)} \, , \label{sol} 
\ee
$(a_1, b_1)$ are integration constants and $b_1$ is removable by time 
rescaling. Note immediately that the range $1/4 < \s < 1$ is excluded to 
retain the signature. There are two special values of $\s$: At $\s=1$, 
there is no solution at all, as is obvious from (\ref{ract}). To our 
knowledge, this is the only gravitational model without a spherical 
metric! For $\s = 1/4$, one finds 
\be 
a(r) = \ln{(\frac{r}{r_{0}})} \, , \qquad b(r) = \frac{1}{r} \, ,
\ee
instead of (\ref{sol}); there is a horizon at $r=r_{0}$. The singularity
at $r=0$ can be seen from 
\be
R_{\m\n\a\b} R^{\m\n\a\b} = \frac{4}{r^4} \Big( 6 - 12 \ln{(\frac{r}{r_{0}})} 
+ 7  \ln^{2}{(\frac{r}{r_{0}})} \Big) \, , \qquad
R = \frac{2}{r^2} \Big( 1 - \ln{(\frac{r}{r_{0}})} \Big) \, .
\ee
Both invariants vanish at infinity, and \( R_{\m\n\a\b} R^{\m\n\a\b} \) is 
positive at all finite $r$. On the other hand, the curvature scalar changes
its sign on the horizon. As $r \to \infty$, both $g_{00}$ and $g_{rr}$ go 
to 0, so this model is unphysical.

The independent curvature invariants for the generic solution, 
$\s < 1/4$ and $1 < \s$, are
\begin{eqnarray}
R_{\m\n\a\b} R^{\m\n\a\b} & = & 
\frac{12 \, r^{-4+2/(\s-1)}}{(1 - \s)^2 (1 - 4 \s)^2}
\Big( 3 \s^2 \, r^{2/(1-\s)} \, (4 - 2 \s + 7 \s^2) \nonumber \\ 
& & \hspace{-0.5in}
+ 6 a_1 \, \s \, r^{(1+4\s)/(1-\s)} \,  
(1 - 3 \s - 3 \s^2 - 4 \s^3) 
+ a_1^2 \, r^{8 \s/(1-\s)} \, (1 - 4\s)^2 (1 + 5 \s^2) \Big) \, , \nonumber \\
R & = & \frac{6 \s}{r^2 (1 - \s) (1 - 4 \s)} 
\Big( -3 \s + (4 \s - 1) \, a_1 \, r^{(1 - 4 \s)/(\s-1)}) \Big) \, .
\end{eqnarray}
For $\s = 0$, one recovers the GR results 
(\( R_{\m\n\a\b} R^{\m\n\a\b} = 12 a_1^{2} / r^6 \,, R = 0 \))
as (\ref{sol}) limits to the Schwarzschild metric for $a_1<0$. 
The spatial behavior of these invariants can be mapped for the allowed 
ranges of $\s$ and compared to GR. Both scalars vanish at infinity, and 
diverge only at the origin for all viable $\s$. For physical, negative $a_1$, 
\( R_{\m\n\a\b} R^{\m\n\a\b} \) is positive at all finite $r$. The
curvature scalar $R$ does vanish, and changes sign, at some finite $r$,
when $0 < \s < 1/4$ [unlike the $\s = 1/4$ case, not at the horizon!]. 
For $\s < 0$ and $1 < \s$, $R<0$ has no sign flip.

The pure Weyl, $\s \to \infty$ end, exhibits slower fall-off than 
Schwarzschild, namely 
\( R_{\m\n\a\b} R^{\m\n\a\b} = 63/(4 r^4) \,, R=-9/(2 r^2) \) for $a_1 = 0$.
Moreover, in the pure Weyl case, one can show that the spacetime is 
conformally flat, \( C_{\m\n\a}~^{\b} = 0 \). That is, $X$ vanishes on
Weyl shell, just as $R$ did on GR shell at $\s=0$. Note that our ``Weyl''
actions are {\it not} conformally invariant (unlike the $C^2$ action),
not being of degree zero in the metric. Correspondingly, our conformal
factor is fixed rather than arbitrary \footnote{There is a slight
formal caveat here, which we have not pursued in detail, given the
assurance provided by \cite{P}. Stated succinctly, if one varies the
full action \( \int d^4 x \sqrt{-g} \, \sqrt{ \mbox{tr} \, C^2 } \),
the resulting field equation contains a term of the form
\( \nabla \nabla (C / \sqrt{ \mbox{tr} \, C^2 }) \), which seems to
have the form $0/0$ in $C=0$ spaces (the other terms vanish there). 
However, we believe that performing the differentiations and then
going to our ansatz will result in the equations we have found above.
In any case, the Einstein term acts as a regulator while the 
$\s \to \infty$ limit is taken.}. 

Since the asymptotic behaviors of $g_{00}$ and $g_{rr}$ differ, the
equivalence principle is violated. (This is a bit loose since we have
defined neither inertial nor gravitational mass. To see the difficulty
of defining a ``mass'' in this theory, one can consider the small $\s$
limit, say to order $\s^2$. At this order $g_{00}$ and $g_{rr}$ have $\ln{r}$
terms in addition to GR's $1/r$ parts.) However, it must be
admitted that there is no qualitative surprise here, such as loss of
origin singularity or horizons, apart from the $\s=1$ loss of solution.
 
\section{\label{maxcos} Generalizations and other $D$}

In this section, we extend the above exercise to include a cosmological 
constant, Maxwell matter and allow general dimension:
\be 
I = \frac{1}{2} \, \int d^D x \, \sqrt{-g} \, \Big( R + \Lambda + \b_{n} \, 
| \mbox{tr} \, C^n|^{1/n} - \frac{1}{4} F_{\m\n} \, F^{\m\n} \Big) \, . 
\label{actD}
\ee
The form of the metric is unchanged,
\be
ds^2 = - a(r) \, b^2(r) \, dt^2 + \frac{dr^2}{a(r)} + r^2 \, d\O_{D-2} \; .
\ee
Its curvature scalar reads
\be 
R = -\frac{1}{r^2} \Big( (D-2)(D-3)(a-1) + 2 (D-2) r a^{\prime} 
+ r^2 a^{\prime\prime} \Big) - \frac{1}{r b} (3 r a^{\prime} b^{\prime}
+ 2 a ((D-2) b^{\prime} + r b^{\prime\prime}) ) \, . 
\ee
In $D$-dimensions, the expressions analogous to those in $D=4$ become \cite{DR}
\be 
\mbox{tr} \, C^n = \Big( -\frac{D-3}{D-1} \Big)^{n} \; 
\Big[ 1- 2 (2-D)^{1-n} + \Big( \frac{2}{(D-2)(D-3)}\Big)^{n-1} 
\Big] \, X^{n} \, , 
\ee
where now
\be 
\s \equiv \b \, \frac{D-3}{D-1} \; \Big| 1- 2 (2-D)^{1-n} 
+ \Big( \frac{2}{(D-2)(D-3)}\Big)^{n-1} \Big|^{1/n} \, .
\ee
Including the Maxwell term, with \( A_{\m} = (A_0(r), 0, \dots, 0) \),
reduces the total action (\ref{actD}) to
\begin{eqnarray}
\frac{1}{2} \, \int_{0}^{\infty} dr \Big[ (D-2)(1-\s) a \, b^{\prime} 
\, r^{D-3} + \s (D-2) (D-1) a \, b \, r^{D-4} \nonumber \\
+ [(D-2)(D-3) - 2 \s] b \, r^{D-4} + \Lambda \, b \, r^{D-2}
+ \frac{r^{D-2}}{2 b} \, (A_{0}^{\prime})^2 \Big]
\, . \label{ractD}
\end{eqnarray}

For generic values of the parameters in the action, the solution of the
field equations arising from (\ref{ractD}) reads
\begin{eqnarray}
a(r) & = & \frac{(D-3)(D-2) - 2 \s}{(D-2)((D-3) - 2 \s (D-2))} 
+ a_1 \, r^{(2 \s (D-2) - (D-3))/(1 - \s)} \nonumber \\
& & + \frac{q^2 \, r^{6-2D}}{2 (D-2) (2 \s + (D-3))} 
+ \frac{\Lambda \, r^2}{(1 - 2 \s) (D-2)(D-1)} \, , \label{asol} \\
b(r) & = & b_1 \, r^{\s (1-D)/ (1 - \s)} \, , \quad
A_{0}^{\prime} = q \, b(r) \, r^{2-D} \, , \label{bsol}
\end{eqnarray}
where $a_1$, $b_1$ and $q$ are integration constants.

Just as in the $D=4$ case, there are some special values for $\s$. The 
$\s=1$ case still excludes a solution. For $\Lambda \neq 0$, there is
one additional point: For $\s = 1 /2$, 
\be 
a(r) = - \frac{D^2 - 5 D + 5}{D-2} + 
\frac{2 \Lambda}{(D-2)} r^2 \ln{a_1 \, r} 
+ \frac{q^2}{2 (D-2)^2} \, r^{6-2D} \, , \quad
b(r) = b_1 \, r^{1-D} \, . 
\ee
For $q \neq 0$, $\s = - (D-3)/2$ is a special point for which
\be 
a(r) = \frac{1}{D-2} + \frac{\Lambda \, r^2}{(D-1)(D-2)^2} 
- \frac{q^2}{(D-2) (D-1)} \, r^{6-2D} \ln{a_1 \, r} \, , \quad
b(r) = b_1 \, r^{D-3} \, . 
\ee
The remaining special cases are \( \s = (D-3)/(2 (D-2)) \) for which
\be 
a(r) = \frac{\Lambda \, r^2}{(D-1)} + \frac{q^2}{2 (D-3) (D-1)} 
\, r^{6-2D} + \frac{2 (D-3)^2}{D-2} \, \ln{a_1 \, r} \, , \quad
b(r) = b_1 \, r^{3-D} \, , 
\ee
and finally $\Lambda=q=0$ reduce (\ref{asol}), (\ref{bsol}) to the pure
``Weyl case''
\be 
a(r) = \frac{1}{(D-2)^2} + a_1 \, r^{4-2D} \, , \quad
b(r) = b_1 \, r^{D-1} \, . 
\ee

A fair conclusion of this Section's various generalizations to include
Maxwell matter, a cosmological term and arbitrary dimension is that they
have not led to any substantial new surprises beyond the original model's.

\section{\label{summ} Summary:}
We have investigated the class of gravitational additions to GR that
preserves GR's derivative order in the spherically symmetric problem:
it is uniquely found to be \( (\mbox{tr} \, C^n)^{1/n} \).
This choice was made -- apart from its easy solubility -- to see how
radically a physically natural extension of GR alters the Schwarzschild
metric's properties. Our findings for the simplest $D=4$ story led to
rather strange metrics, which however retain the former's qualitative
properties: horizon and origin singularity, apart from the complete loss
of a solution at $\s=1$. These properties are essentially maintained by
our further generalizations.

\begin{acknowledgments}
S.D. was supported in part by NSF grant PHY04-01667, and thanks Dr. G.
Conrad for a stimulating contribution. The work of {\"O}.S. and B.T. is 
partially supported by the Scientific and Technological Research Council 
of Turkey (T{\"U}B\.{I}TAK). B.T. is also partially supported by the 
``Young Investigator Fellowship" of the Turkish Academy of Sciences 
(T{\"U}BA) and by the T{\"U}B\.{I}TAK Kariyer Grant 104T177.
\end{acknowledgments}

\appendix*

\section{Gauss-Bonnet augmented model}
Here we add a Gauss-Bonnet term to (\ref{actD}) which, of course, only 
contributes for $D \ge 5$:
\begin{eqnarray}
\frac{\g \, \sqrt{-g}}{\O_{D-2}} (R^2 - 4 R_{\m\n}^{2} + R_{\m\n\a\b}^{2}) 
& = & \g \Big[ (D-2) (D-3) r^{D-4} \Big( 2 a a^{\prime} b + 4 a^2 b^{\prime}
+(D-4) a^2 b/r \nonumber \\
& & - 2 (D-4) ab/r - 2 a^{\prime} b - 4 a b^{\prime} + (D-4) b/r \Big) 
\Big]^{\prime} \nonumber \\
& & - \g (D-4) (D-3) (D-2) r^{D-5} (a-1)^{2} \, b^{\prime} \, .
\end{eqnarray}
The bracketed term is a total divergence, so the overall reduced action
is still of first derivative order
\begin{eqnarray}
\frac{1}{2} \, \int_{0}^{\infty} dr \Big[ (D-2)(1-\s) a \, b^{\prime} 
\, r^{D-3} + \s (D-2) (D-1) a \, b \, r^{D-4} \nonumber \\
+ [(D-2)(D-3) - 2 \s] b \, r^{D-4} + \Lambda \, b \, r^{D-2}
+ \frac{r^{D-2}}{2 b} \, (A_{0}^{\prime})^2 
- \lambda \, r^{D-5} \, (a-1)^{2} \, b^{\prime} \Big] \, ,
\end{eqnarray}
here \( \lambda \equiv \g (D-4) (D-3) (D-2) \). Without the ``Weyl'' term, 
the general solution is well known \cite{BD}, \cite{DT}. As a specific 
example, consider $D=5$ with $\Lambda=0$ and $A_{0}=0$. For the generic 
values of the parameter $\s$ (including the $\s=1$ case, which is here 
resuscitated!), one can obtain a solution: The function $b(r)$ can be 
solved in terms of $a(r)$; however the solution for $a(r)$ itself is best 
not displayed. There is one special case $\s = 1/3$ for which the solution is 
``somewhat'' presentable:
\be 
a(r) = 1 + \frac{1}{\lambda} \, r^2 - \frac{4}{3} \Big( 1 + a_1
+ W( - e^{-1 + 3 r^2 /(4 \lambda)} ) \Big) \, , \quad
b(r) = e^{(b_1 - 3 a(r))/4} \, , 
\ee
where $W(z)$ denotes the Lambert $W$ function, i.e. \( z = W(z) \, e^{W(z)} \) 
and $W(z)$ is single valued for real $z \geq -1/e$ with 
$W(z) \geq -1$ demanded.


\begin{thebibliography}{99}

\bibitem{W} H. Weyl, ``{\it Space-Time-Matter}'', New York: Dover, (1951).

\bibitem{P} R.S. Palais, {\it Comm. Math. Phys.} {\bf 69}, 19 (1979).

\bibitem{DT} S. Deser and B. Tekin,
  Class.\ Quant.\ Grav.\ {\bf 20}, 4877 (2003) [arXiv:gr-qc/0306114].

\bibitem{DR} S. Deser and A.V. Ryzhov,
  Class.\ Quant.\ Grav.\  {\bf 22}, 3315 (2005) [arXiv:gr-qc/0505039].

\bibitem{DF} S. Deser and J. Franklin,
  Am.\ J.\ Phys.\ {\bf 73}, 261 (2005) [arXiv:gr-qc/0408067].

\bibitem{BD} D.G. Boulware and S. Deser,
  Phys.\ Rev.\ Lett.\  {\bf 55}, 2656 (1985).

\end{thebibliography}
\end{document}